\documentstyle[12pt,aasms4] {article}
\newcommand{\beq}{\begin{equation}}
\newcommand{\eeq}{\end{equation}}
\newcommand{\beqn}{\begin{eqnarray}}
\newcommand{\eeqn}{\end{eqnarray}}

\newcommand{\pa}{\partial}
\newcommand{\no}{\noindent}

\newcommand{\non}{\nonumber}

\begin{document}
\title{ On the Chermnykh-Like Problems:\\
        II. The Equilibrium Points }

\author{Li-Chin Yeh$^{1}$ and
Ing-Guey Jiang$^{2}$}

\affil{
{$^{1}$ Department of Applied Mathematics,}\\
{ National Hsinchu University of Education, Hsin-Chu, Taiwan} \\
{$^{2}$ Institute of Astronomy,}\\
{ National Central University, Chung-Li, Taiwan} 
}

\authoremail{jiang@astro.ncu.edu.tw}

\begin{abstract} 
Motivated by Papadakis (2005a, 2005b), 
we study a Chermnykh-like problem, in which 
an additional gravitational potential from the belt is included. 
In addition to the
usual five equilibrium points  
(three collinear and two triangular points), 
there are some new equilibrium points for 
this system.
We studied the conditions for the existence of these new equilibrium
points both analytically and numerically. 
\end{abstract}

\newpage
\section{Introduction}

The interest in the Chermnykh's problem (Chermnykh 1987)  
has been renewed by K.E. Papadakis
recently. For example, Papadakis (2004) studied the symmetric motions
near the three collinear equilibrium points in three dimensional space.
Papadakis (2005a) numerically investigated the equilibrium points
and the zero-velocity curves under the assumption that the mass 
parameter is fixed but the angular velocity parameter is allowed to vary
continuously. The stability of the periodic orbits are determined and
the network of the orbital families is explored. 

Moreover, Papadakis (2005b) numerically studied the asymmetric periodic 
orbits near the triangular equilibrium points 
under the assumption that the angular velocity varies and for the 
Sun-Jupiter mass distribution. Remarkably, that paper provides
the analytic determination of the initial conditions of the long- 
and short-period Trojan families around the equilibrium points.

On the other hand, astronomers claimed that 
it is common to have
circumbinary discs for binary systems and asteroid belts
for planetary systems. Some studies showed that these belt-like 
structures shall influence the mathematical properties of 
these dynamical systems. The number and locations of equilibrium
points and also topology of solution curves might become very different
when the influence from the belt is considered.
For example, 
Jiang \& Yeh (2003, 2004) considered the influence from the belt 
for planetary systems and found that the probability to have equilibrium points
around the inner part of the belt is larger than the one near the outer
part. Their results can be used to explain the observational configuration
of Kuiper Belt Objects of the outer Solar System.

Jiang \& Yeh (2006) have studied a Chermnykh-like problem in which the mass
parameter $\mu$ is set to be 0.5.
We now study the systems with any 
possible mass-ratio of central binaries both analytically and numerically. 
We will focus on 
conditions with which the
new equilibrium points can exist for two different density models
of the belts.
We can then understand the locations
of these new equilibrium points for given parameters. 
We present our models in Section 2,
the results for the existence of new equilibrium points are in Section 3.
Section 4 is about the stability analysis
and Section 5
concludes the paper.
 
\section{The Models}

We consider the motion of a test particle influenced by 
the gravitational force from the central binary and the circumbinary belt.
We assume that two masses of the central binary 
are $m_1$ and $m_2$ and choose
the unit of mass to make $G(m_1+m_2)=1$. If we define that 
$${\mu}=\frac{m_2}{m_1+m_2},$$
then 
the two masses are
$\mu_1=Gm_1=1-{\mu}$ and $\mu_2=Gm_2={\mu}$.
The separation of central binary objects 
is set to be unity. In this paper, we assume 
that $\mu \le 0.5$. The location of two masses are always at 
$(1-\mu, 0, 0)$ and  $(-\mu, 0, 0)$. 

The equation of motion is 
(Jiang \& Yeh 2006)
\beq\left\{
\begin{array}{ll}
&\frac{dx}{dt}=u \\
&\frac{dy}{dt}=v  \\
&\frac{du}{dt}=2nv-\frac{\pa U^{\ast}}{\pa x}-\frac{\pa V}{\pa x}  \\
& \frac{dv}{dt}=-2nu-\frac{\pa U^{\ast}}{\pa y}-\frac{\pa V}{\pa y},
\end{array} \right. \label{eq:3body1} 
\eeq
where $n$ is the central binary's angular velocity (Please note that we
keep $n$ as a parameter from here until Property 3.2 but set it 
to be 1 as a simplification starting from Eq.(20), i.e, Section 3.1 Model A.),
\beq
U^{\ast}=-\frac{n^2}{2}(x^2+y^2)-\frac{1-\mu}{r_1}-\frac{\mu}{r_2},
\label{eq:u_ast}
\eeq
\beq
r_1=\sqrt{(x+\mu)^2+y^2},\label{eq:r1}
\eeq
\beq 
 r_2=\sqrt{(x+\mu-1)^2+y^2}\label{eq:r2}
\eeq
and 
$V$ is the potential from the belt. We assume that $V$ is radial symmetric, 
so $V$ depends on the radial distance $r$, where $r={\sqrt {x^2+y^2}}$. Hence,
\beq\left\{
\begin{array} {ll}
&\frac{\pa V}{\pa x}= \frac{x}{r}\frac{\pa V}{\pa r}  \\
&\frac{\pa V}{\pa y}= \frac{y}{r}\frac{\pa V}{\pa r}, 
\end{array} \right.\label{eq:pav}
\eeq

We substitute Eq. (\ref{eq:u_ast}) and Eq. (\ref{eq:pav}) into 
Eq.(\ref{eq:3body1}) and have the following system:

\beq \left\{
\begin{array}{ll}
&  \frac{dx}{dt}=u  \\
& \frac{dy}{dt}=v  \\
& \frac{du}{dt}=2nv + n^2 x-\frac{(1-\mu)(x+\mu)}{r_1^3}-\frac{\mu(x+\mu-1)}
{r_2^3}-\frac{x}{r}\frac{\pa V}{\pa r}  \\
& \frac{d v}{dt}=-2nu+ n^2 y-\frac{y(1-\mu)}{r_1^3}-\frac{y\mu}{r_2^3}
-\frac{y}{r}\frac{\pa V}{\pa r} ,  
\end{array}  \right. \label{eq:3body2}
\eeq
where $r_1$ and $r_2$ are defined in Eqs.(\ref{eq:r1})-(\ref{eq:r2}). 
We can consider any density 
profile of the belt by giving a formula of $V$ in Eq.
(\ref{eq:3body2}) as long as $V$ is radial symmetric.  In this paper, we
study two different models of the belt. 

\subsection {Model A: Miyamoto-Nagai Profile}
In Model A, we introduce the belt potential, from Miyamoto \& Nagai (1975),
which is 
\beq
V(r,z)=-\frac{M_b}{\sqrt{r^2+(a+\sqrt{z^2+b^2})^2}},\label{eq:va1}
\eeq
where $M_b$ is the total mass of the belt and $r^2=x^2+y^2$. 
In this formula, $a$ and 
$b$ are parameters which determine the density profile of the belt.
The parameter $a$ controls the flatness of the profile and can be called 
``flatness parameter''. The parameter $b$ controls the size of the core
of the density profile and can be called ``core parameter''.
When $a=b=0$, the potential equals to the one by a point mass. 
In general, the density distribution corresponding to the above $V(r,z)$ 
in Eq.(\ref{eq:va1}) is, as in Miyamoto \& Nagai (1975),    
\beq
\rho(r,z)=\left(\frac{b^2M_b}{4\pi}\right)\frac{ar^2+(a+3\sqrt{z^2+b^2})
(a+\sqrt{z^2+b^2})^2}{[r^2+(a+\sqrt{z^2+b^2})^2]^{5/2}(z^2+b^2)^{3/2}}.
\eeq
If we define $T\equiv a+b$, from Eq. (\ref{eq:va1}) we have
\beq
\frac{\pa V}{\pa r}(r,0)=\frac{M_b r}{(r^2+(a+b)^2)^{3/2}}=
\frac{M_b r}{(r^2+T^2)^{3/2}},\label{eq:va2}
\eeq
where we set $z=0$ since we only consider the orbits on the $x-y$ plane.
We substitute Eq. (\ref{eq:va2}) into Eq. (\ref{eq:3body2}) and have
the following system:
\beq \left\{
\begin{array}{ll}
&  \frac{dx}{dt}=u  \\
& \frac{dy}{dt}=v  \\
& \frac{du}{dt}=2nv + n^2 x-\frac{(1-\mu)(x+\mu)}{r_1^3}-\frac{\mu(x+\mu-1)}
{r_2^3}-\frac{M_b x}{(r^2+T^2)^{3/2}}  \\
& \frac{d v}{dt}=-2nu+ n^2 y-\frac{y(1-\mu)}{r_1^3}-\frac{y\mu}{ r_2^3}
-\frac{M_b y}{(r^2+T^2)^{3/2}}.   
\end{array}  \right. \label{eq:ma3body2}
\eeq
We can see that neither the core parameter or the flatness parameter
appear in the equations of motion. The dynamics of the system only 
depends on the summation of $a$ and $b$, i.e. $T$. 
 
\subsection {Model B: Power-Law Profile }
The belt is a annulus with inner radius $r_{i}$ and
outer radius $r_{o}$, where  $r_{i}$ and $r_{o}$ are assumed to be 
constants. The density profile of the  belt is 
\beq
\rho(r)=\left\{\begin{array}{ll}
0  & {\rm  when}\,\,  r < r_{i} \,\,{\rm or}\,\,  r > r_{o},\\
 \frac{c}{r^p}\left\{\cos\left[ \frac{\pi}{2} \frac{(r-r_1)}{(r_{i} - r_1)} 
\right]\right\}^2 & 
{\rm  when} \,\,   r_{i} < r < r_1,  \\
 \frac{c}{r^p} & {\rm when} \,\,   r_1 < r < r_2,  \\
\frac{c}{r^p}\left\{\cos\left[ \frac{\pi}{2} \frac{(r-r_2)}{(r_{o}-r_2)} 
\right]\right\}^2 & 
{\rm  when} \,\,   r_2 < r < r_{o},  \\
\end{array}\right.
\eeq
 where 
$r=\sqrt{x^2+y^2}$, $c$ is a constant 
completely determined by the total mass of the belt and 
we set $p=2$ (Lizano $\&$ Shu 1989), $r_1=r_i+0.1$, $r_2=r_1+0.8$, and 
$r_o=r_2+0.1$ for all numerical results. Hence, 
for $p=2$, the total mass of the belt is 
\beqn
M_{b}&=&\int^{2\pi}_{0}\int^{r_o}_{r_i}\rho(r')r'dr'd\phi =2\pi c
\left\{\int^{r_1}_{r_i}\frac{1}{r'}\left\{\cos\left[ \frac{\pi}{2} 
\frac{(r-r_1)}{(r_{i} - r_1)} \right]\right\}^2 dr'+
\ln (r_2/r_1) \right. \non \\
& &\left. +\int^{r_o}_{r_2}\frac{1}{r'}\left\{\cos\left[ \frac{\pi}{2} 
\frac{(r-r_2)}{(r_{o}-r_2)} \right]\right\}^2 dr'\right\}. 
\eeqn
The gravitational force $f_b$ from the belt is (Jiang $\&$ Yeh 2004) 
\beq
f_b(r)=-\frac{\pa V}{\pa r}= 
-2\int^{r_o}_{r_i}\frac{\rho(r')r'}{r}\left[\frac{E(\xi)}{r-r'}+
\frac{F(\xi)}{r+r'}\right] dr',\label{eq:fb}
\eeq
where $\xi=2\sqrt{r r'}/{(r+r')}$, $F$ and $E$ are elliptic 
integrals of the first kind and the second kind.
We substitute  Eq. (\ref{eq:fb}) into Eq.(\ref{eq:3body2}) and have 
the following system:
\beq \left\{
\begin{array}{ll}
&  \frac{dx}{dt}=u  \\
& \frac{dy}{dt}=v  \\
& \frac{du}{dt}=2nv + n^2 x-\frac{(1-\mu)(x+\mu)}{r_1^3}-\frac{\mu(x+\mu-1)}
{r_2^3}-\frac{2x}{r^2}  \int^{r_o}_{r_i}\rho(r')r'
\left[\frac{E}{r-r'}+\frac{F}{r+r'}\right]dr' \\
& \frac{d v}{dt}=-2nu+ n^2 y-\frac{y(1-\mu)}{r_1^3}-\frac{y\mu}{r_2^3}
-\frac{2y}{r^2}  \int^{r_o}_{r_i}\rho(r')r'\left[\frac{E}{r-r'}
+\frac{F}{r+r'}\right]dr'.   
\end{array}  \right. \label{eq:mb3body2}
\eeq

\section{The New Equilibrium Points}

It is well-known that there are five equilibrium points, i.e. 
three collinear and two triangular points, for the
classical restricted three-body problem.
When there are more than five equilibrium points for the system we 
study here, we claim that
the new equilibrium points exist.

In general, for System 
(\ref{eq:3body2}), equilibrium points $(x_e,y_e)$ satisfies $f(x_e,y_e)=0$ and
$g(x_e,y_e)=0$, where 
\beqn
&&f(x,y)=n^2 x-\frac{(1-\mu)(x+\mu)}{ r_1^3}-\frac{\mu(x+\mu-1)}
{r_2^3}-\frac{x}{r}\frac{\pa V}{\pa r},  \label{eq:gf2}\\
& & g(x,y)=n^2 y-\frac{y(1-\mu)}{r_1^3}-\frac{y\mu}{r_2^3}
-\frac{y}{r}\frac{\pa V}{\pa r}. \label{eq:gg2}
\eeqn

For convenience, we define 
\beq
h(y)\equiv n^2 -\left[\frac{1}{4}+y^2\right]^{-3/2}
-\frac{1}{r}\frac{\pa V}{\pa r}\biggm|_{(\frac{1}{2}-\mu,y)} 
\label{eq:h}
\eeq 
and
\beq
k(x)\equiv n^2  x-\frac{(1-\mu)(x+\mu)}{|x+\mu|^3}-\frac{\mu(x+\mu-1)}
{|x+\mu-1|^3} - \frac{x}{r}\frac{\pa V}{\pa r}\biggm|_{(x,0)}. \label{eq:k}
\eeq
We have the following properties:

{\bf  Property {3.1}}\\ 
 {\it  If $(x_e,y_e)$ is an equilibrium point of System (\ref{eq:3body2}), 
then we have\\
 either (1) $x_e=1/2-\mu$ and $y_e$ satisfies $h(y)=0$  \\
or (2) $x_e$ satisfies $k(x)=0$ and $y_e = 0$.
}

{\it Proof:} Suppose that $(x_e,y_e)$ is an equilibrium point, thus
it satisfies $f(x,y)=0$ and  $g(x,y)=0$.
From Eq.(\ref{eq:gg2}), we have
$$y_e\left[n^2 -\frac{1-\mu}{r_1^3}-\frac{\mu}{r_2^3}
-\frac{1}{r}\frac{\pa V}{\pa r}\right]_{(x_e,y_e)}=0.$$
Hence, $y_e=0$ or $n^2 -\frac{1-\mu}{r_1^3}-\frac{\mu}{r_2^3}
-\frac{1}{r}\frac{\pa V}{\pa r}\biggm|_{(x_e,y_e)}=0$.  
We discuss these two cases separately:

(I) $n^2 -\frac{1-\mu}{r_1^3}-\frac{\mu}{r_2^3}
-\frac{1}{r}\frac{\pa V}{\pa r}\biggm|_{(x_e,y_e)}=0$:\\ 
Since $(x_e,y_e)$ is an equilibrium point, that is $f(x_e,y_e)=0$, we have
\beq
 0=f(x_e,y_e)= x_e\left[n^2 -\frac{1-\mu}{r_1^3}-\frac{\mu}{r_2^3}
-\frac{1}{r}\frac{\pa V}{\pa r}\right]-\frac{(1-\mu)\mu}{r_1^3}
+\frac{(1-\mu)\mu}{r_2^3}
=(1-\mu)\mu\left[-\frac{1}{r_1^3}+\frac{1}{r_2^3},\right]. \label{eq:con2} 
\eeq
Thus,  $r_1=r_2$, i.e.
$(x_e+\mu)^2+y_e^2=(x_e+\mu-1)^2+y_e^2$. Hence $x_e=1/2-\mu$.
We have  $r_1=r_2=\sqrt{1/4+y_e^2}$ and thus, 
$n^2-\frac{1-\mu}{r_1^3}-\frac{\mu}{r_2^3} =
n^2-\left[\frac{1}{4}+y^2\right]^{-3/2} $.
Therefore,  
 $x_e=1/2-\mu$ and $y_e$ satisfies 
$h(y)=0$ for 
the equilibrium point $(x_e,y_e)$.

(II) $y_e=0$:\\
 $f(x_e,y_e)= f(x_e,0)= k(x_e)= 0$ for 
the equilibrium point $(x_e,y_e)$.  Thus, 
$x_e$ satisfies $k(x)=0$ and $y_e = 0$. $\Box$

{\bf  Property {3.2}} \\
\no {\it (A) If  $y_e$ satisfies  $h(y)=0$, then 
$(1/2-\mu,y_e)$  is 
the equilibrium point of System (\ref{eq:3body2}).\\
\no (B) If $x_e$ satisfies $k(x)=0$, then 
$(x_e,0)$  is  the equilibrium point of System (\ref{eq:3body2}).}


{\it Proof of (A):}
Suppose that $y_e$ is one of the roots of $h(y)$, i.e. $h(y_e)=0$ and 
we set $x_e=1/2-\mu$.
Because $g(1/2-\mu,y_e) = y_e h(y_e) =0$ and
$f(1/2-\mu,y_e) = \left(\frac{1}{2}-\mu\right) h(y_e) =0$, 
$(x_e,y_e)$ is the equilibrium point of System (\ref{eq:3body2}). $\Box$

{\it Proof of (B):}
Suppose that $x_e$ is one of the roots of $k(x)$, i.e. $k(x_e)=0$ 
and we set $y_e=0$. Since $y_e=0$, it is trivial that $g(x_e,y_e) =0$.
Because $k(x_e)=0$, $f(x_e,0)=k(x_e)=0$. 
Thus, $(x_e,y_e)$ is the equilibrium point
of System (\ref{eq:3body2}). $\Box$ 


Because of  the above two properties, in stead of searching the roots
for two variable functions $f(x,y)$ and $g(x,y)$ 
to determine all equilibrium points
on the $x-y$ plane, 
we only need to find the roots of one variable functions $h(y)$ and $k(x)$
to get all the equilibrium points of System (\ref{eq:3body2}).
 
\subsection{Model A}
We now study the equilibrium points when $V$ is Miyamoto-Nagai profile.
From Property 3.1, equilibrium points should be either on 
the $x$-axis or on the line: $x=1/2-\mu$ of $x-y$ plane. 
As a simplification, please note that we set $n=1$ hereafter.   
For equilibrium points  
on the $x$-axis, i.e. $(x_e,0)$, 
we have
\beq
k(x_e)=x_e-\frac{(1-\mu)(x_e+\mu)}{|x_e+\mu|^3}-\frac{\mu(x_e+\mu-1)}
{|x_e+\mu-1|^3}-\frac{x_eM_b}{(x_e^2+T^2)^{3/2}}=0.\label{eq:fxe}
\eeq

On the other hand, for those 
equilibrium points on the line: $x=1/2-\mu$
(i.e. $(1/2-\mu,y_e)$), and thus
$r=\sqrt{(1/2-\mu)^2+y_e^2}$,
we have 
\beq
h(y_e)= 1 - \left[\frac{1}{4}+y_e^2\right]^{-3/2}
-M_b\left[\left(\frac{1}{2}-\mu\right)^2+y_e^2+T^2\right]^{-3/2}=0.
\label{eq:gxe}
\eeq


{\bf  Property {3.3}}\\
 {\it There is one and only one 
$\bar{y_1}>0$ such that $h(\bar{y_1})=0$, and only one 
$\bar{y_2}<0$ such that $h(\bar{y_2})=0$. That is, 
there are two equilibrium points on the line: $x=1/2-\mu$
for System (\ref{eq:ma3body2}).}

{\it Proof: }  
We define
\beq
P_1(y)=\left[\frac{1}{4}+y^2\right]^{-3/2}-1 \label{eq:q1}
\eeq
 \beq
{\rm and} \qquad Q_1(y)
=-M_b\left[\left(\frac{1}{2}-\mu\right)^2+y^2+T^2\right]^{-3/2}, 
\label{eq:p1}
\eeq 
so from Eq.(\ref{eq:gxe}), we have 
\beq
h(y)
=1-\left[\frac{1}{4}+y^2\right]^{-3/2}-M_b\left[\left(\frac{1}{2}-\mu
\right)^2+y^2+T^2\right]^{-3/2} 
  \equiv -P_1(y)+Q_1(y)=0.\non 
\eeq

From Eqs. (\ref{eq:q1})-(\ref{eq:p1}), we have
$$Q'_1(y)=3yM_b\left[\left(\frac{1}{2}-\mu\right)^2+y^2+T^2\right]^{-5/2}
{\rm and} \quad
P'_1(y)= -3y\left[\frac{1}{4}+y^2\right]^{-5/2}. $$
Since $P'_1(y)<0$ for $y>0$, $P_1(y)$ is a monotonically 
decreasing and continuous 
function
for any $y\in (0, \infty)$. $Q'_1(y)>0$ for $y>0$, thus 
$Q_1(y)$ is a monotonically increasing and continuous 
function for any $y\in (0, \infty)$.
Because $P_1(0)>0>Q_1(0)$ and  
$\lim_{y\to \infty}P_1(y)< \lim_{y\to \infty}Q_1(y)$,
there exists 
a unique point $\bar{y_1}>0$ such that $P_1(\bar{y_1})=Q_1(\bar{y_1})$, 
that is $h(\bar{y_1})=0$.

By the similar method, we find that there is  
a unique point $\bar{y_2}<0$ such that 
$h(\bar{y_2})=0$. Hence, there are two equilibrium points on the 
line: $x=1/2-\mu$. $\Box$

In the classical restricted three-body problem, there
are three collinear points and two triangular points, i.e. the points 
on the line: $x=1/2-\mu$.
Therefore, Property 3.3
shows that there is no new equilibrium points on the line:
$x=1/2-\mu$ for Model A. If there is any new equilibrium points
for this model, it must be on the $x$-axis.
 
Let 
\beq
P_2(x)=\frac{(1-\mu)(x+\mu)}{|x+\mu|^3}+\frac{\mu(x+\mu-1)}
{|x+\mu-1|^3}-x, \label{eq:p2}
\eeq
and thus,
\beq
P_2(x) =\left\{
\begin{array} {ll}
&\frac{1-\mu}{(x+\mu)^2}+
\frac{\mu}{(x+\mu-1)^2}-x, \quad {\rm for} \,\, x>1-\mu, \\
& \frac{(1-\mu)}{(x+\mu)^2}-
\frac{\mu}{(x+\mu-1)^2}-x, \quad {\rm for} \,\, -\mu<x<1-\mu,\\
& -\frac{1-\mu}{(x+\mu)^2}-
\frac{\mu}{(x+\mu-1)^2}-x, \quad {\rm for} \,\, x<-\mu.
\end{array}\right. \label{eq:p2_3}
\eeq
We also set
\beq
Q_2(x)=-\frac{xM_b}{(x^2+T^2)^{3/2}}. \label{eq:q2}
\eeq 
It is obvious that $P_2(x)$ contains the terms related to the gravitational
forces from the central binary and the centrifugal force. $Q_2(x)$
is simply the term contributed by the belt. By defining these two 
functions, we have both the advantages that the physical meanings 
are clear and some properties about the new equilibrium 
can be proved easily. 

From Eq.(\ref{eq:fxe}), we have $f(x,0)=k(x)= -P_2(x)+Q_2(x)$.
The equilibrium points $(x_e,0)$ satisfy $f(x_e,0)=k(x_e)=0$, that is, 
$P_2(x_e)=Q_2(x_e)$. 
There are two properties for this part of our results:
Property 3.4  is for the case of non-equal mass binaries, i.e. 
$0.5>\mu>0$  and Property 3.5 is  
for the case of equal mass binaries, i.e. 
$\mu=0.5$. 
 In Property 3.4, 
we find that there are three equilibrium points, $(x_i,0)$  for $i=1,2,3$, 
where  $x_1 \in (-\infty,-\mu)$, $x_2 \in (0,1-\mu)$ and 
$x_3 \in (1-\mu,\infty)$. However, the results for the equilibrium points 
which $x-$coordinate is in the region $(-\mu,0)$ are more complicated. 
As we can see in Property 3.4 (C), 
there are two equilibrium points which x-coordinates 
are in the region $(-\mu,0)$, and in Property 3.4 (D), there 
is no equilibrium point with $x-$coordinate in the region $(-\mu,0)$. 
We will see that, in fact,  
Property 3.5 (A) could be combined with Property 3.4 (B).  However,
Property 3.5 (B) studies the same condition as in Property 3.4 (C) 
but the results for that condition are completely different. 
Moreover, There is nothing like Property 3.4 (A) and 3.4 (D) in Property 3.5.
These differences are due to that $P_2(-x)=-P_2(x)$ and
$Q_2(-x)=-Q_2(x)$ when $\mu=1/2$ but the symmetry is broken
when $\mu<1/2$.
We therefore decided to separate Property 3.4 and 3.5.
For convenience, we will 
denote $\mu^{+}$ to represent that $x$ tends to $\mu$ from the right
hand side and $\mu^{-}$ to represent that $x$ tends to $\mu$ from the 
left hand side.

{\bf  Property {3.4}}\\ {\it 
(A) There is an  $x_1>1-\mu$ and  an $x_2\in (0,1-\mu)$ such that 
$k(x_1)=0$ and $k(x_2)=0$. \\
(B) There is an   $x_3 < -\mu$ such that $k(x_3)=0$. \\
(C) If $T<\sqrt{2}\mu$ and $Q_2(-T/\sqrt{2})>P_2(-T/\sqrt{2})$, then 
there are two points in the region $(-\mu,0)$ such that  $k(x)=0$. \\
(D) If $Q_2(-T/\sqrt{2})<P_2(0)$, then there is no point 
in the region $(-\mu,0)$  such that  $k(x)=0$.}

{\it Proof of (A):} From Eqs. (\ref{eq:p2})-(\ref{eq:q2}), we know that if 
$x>1-\mu$, then 
\beq
P_2(x) =
\frac{1-\mu}{(x+\mu)^2}+
\frac{\mu}{(x+\mu-1)^2}-x,
\eeq
and 
$$P'_2(x)=-\frac{2(1-\mu)}{(x+\mu)^3}-\frac{2\mu}{(x+\mu-1)^3}-1.$$
Since $P_2'(x)<0$ for $x>1-\mu$, $P_2$ is a monotonically decreasing function 
for  $x\in (1-\mu,\infty)$.
Because $\lim_{x\to (1-\mu)^{+}}P_2(x)=\infty$, 
$\lim_{x\to \infty}P_2(x)=-\infty$ and $Q_2(1-\mu)<0$,  
$\lim_{x\to\infty}Q_2(x)=0$, we have 
$\lim_{x\to (1-\mu)^{+}} k(x)=\lim_{x\to (1-\mu)^{+}}(-P_2(x)+Q_2(x)) < 0$ and
$\lim_{x\to \infty} k(x)=\lim_{x\to \infty}(-P_2(x)+Q_2(x))> 0$. Therefore, 
there is a point $x_1>1-\mu$ such that $k(x_1)=0$.

In the region$(0,1-\mu)$, $P_2(0)>0$, $\lim_{x\to (1-\mu)^{-}}P_2(x)=-\infty$, 
$Q_2(0)=0$ and $Q_2(1-\mu)<0$, so we have $k(0)=-P_2(0)+Q_2(0)<0$ and 
$\lim_{x\to (1-\mu)^{-}}k(x)=\infty$. Therefore, there is an 
$x_2\in (0,1-\mu)$ such that $k(x_2)=0$. $\Box$


{\it Proof of (B):} If $x<-\mu$, then 
$$P_2(x)=-\frac{1-\mu}{(x+\mu)^2}-\frac{\mu}{(x+\mu-1)^2}-x.$$
Since $P_2'(x)<0$ for $x<-\mu$, $P_2$ is a monotonically 
decreasing function for
$x<-\mu$.

Because $\lim_{x\to -\mu^{-}}P_2(x)=-\infty$,
$\lim_{x\to -\infty}P_2(x)=\infty$ 
and $Q_2(-\mu)>0$, $\lim_{x\to -\infty} Q_2(x)=0$, we have 
$\lim_{x\to -\mu^{-}} k(x) > 0$ and
$\lim_{x\to -\infty} k(x) < 0$. Therefore,
there is a point $x_3<-\mu$ such that $k(x_3)=0$. $\Box$

{\it Proof of (C):} If $x\in (-\mu, 0)$, then 
$$P_2(x)=\frac{1-\mu}{(x+\mu)^2}-\frac{\mu}{(x+\mu-1)^2}-x.$$
If $T<\sqrt{2}\mu$, we have that $-\mu< -T/\sqrt{2}$.  
We then discuss the possible points such that  $k(x)=0$ for $x = -T/\sqrt{2}$,
$x\in (-T/\sqrt{2},0)$ and $x\in (-\mu, -T/\sqrt{2})$  separately.  

If $Q_2(-T/\sqrt{2})>P_2(-T/\sqrt{2})$, then $k(-T/\sqrt{2})=
-P_2(-T/\sqrt{2})+Q_2(-T/\sqrt{2})>0$. Thus, $k(x)\neq 0$ when 
$x = -T/\sqrt{2}$.

Because we set 
$0<\mu<0.5$ in this paper, we have $P_2(0)>0$. 
Since $Q_2(0)=0$, we have $k(0)=-P_2(0)+Q_2(0) < 0$. 
Further, if $Q_2(-T/\sqrt{2})>P_2(-T/\sqrt{2})$,
then $k(-T/\sqrt{2}) >0$. 
Thus, there is an  $x_4\in (-T/\sqrt{2},0)$ such that 
$k(x_4)=0$. 

On the other hand, 
since $P'_2(x)<0$, $Q'_2(x)>0$ for $x\in (-\mu, -T/\sqrt{2})$ ,
$k'(x)>0$ for  $x\in (-\mu, -T/\sqrt{2})$. Thus, $k(x)$ is a 
monotonically increasing function. Because
 $\lim_{x\to -\mu^{+}} P_2(x)=\infty$ and $Q_2(-\mu)>0$, we have
$\lim_{x\to -\mu^{+}} k(x) <0$.
If 
$Q_2(-T/\sqrt{2})>P_2(-T/\sqrt{2})$, then $k(-T/\sqrt{2}) >0$.
Therefore,  there is a unique point
$x_5\in (-\mu,-T/\sqrt{2})$ such that 
$k(x_5)=0$. $\Box$

{\it Proof of (D): }  Since
$P'_2(x)<0$ for $x\in (-\mu, 0)$, $P_2$ is a monotonically decreasing function 
and has a minimum value $P_2(0)$ 
in the region $(-\mu,0)$.
It is easy to show that $Q_2(-T/\sqrt{2})$ is 
the  maximum 
value for $Q_2(x)$ in the region $(-\mu,0)$.
If $Q_2(-T/\sqrt{2})<P_2(0)$, we have $P_2(x) > Q_2(x)$ 
and  thus $k(x)< 0$ in the region $(-\mu,0)$. Therefore,
there is no point such that $k(x)=0$ in the region $(-\mu,0)$. $\Box$

In Property 3.5, we study the $\mu=1/2$ case. Since $P_2(-x)=-P_2(x)$ and
$Q_2(-x)=-Q_2(x)$, 
we only need to discuss $(-\infty,0)$ region.

{\bf  Property {3.5}}\\ {\it 
(A)  There is an  $x_1<-1/2$ such that 
$k(x_1)=0$. \\
(B) If $T<1/\sqrt{2}$ and
$Q_2(-T/\sqrt{2})>P_2(-T/\sqrt{2})$, then  there 
is a point in the region $(-1/2, -T/\sqrt{2} )$ such that $k(x)=0$. }

{\it Proof :} (A) can be proved easily by the 
the same method as the proof for 
Property 3.4 (B).
In the following, we would like to prove (B). Since 
$Q_2(-T/\sqrt{2})>P_2(-T/\sqrt{2})$, $k(-T/\sqrt{2})>0$.
Since
 $Q_2(-1/2)$  is finite and 
$\lim_{x\to (-1/2)^{+}}P_2(x)=\infty$, $\lim_{x\to (-1/2)^{+}} k(x)<0$.
Thus, there is a point
$x^{\ast}\in (-1/2,-T/\sqrt{2})$ such that 
$k(x^{\ast})=0$. $\Box$

From Property 3.5, if $T<1/\sqrt{2}$ and
$Q_2(-T/\sqrt{2})>P_2(-T/\sqrt{2})$,  
we could have two points such that $k(x)=0$ in  
the region $(-\infty, 0)$. By the property of 
symmetry, we will have another 
two roots in the region $(0,\infty)$. Since $k(0)=0$, there are at least 
five points such that $k(x)=0$ for the equal mass case.
We know that there are only three equilibrium points on the $x$-axis 
for the classical restricted
three-body problem, thus new equilibrium points exist 
for this case.
From Property 3.3, 3.4 and Property 3.5, we have the following corollary to 
analytically 
determine the area which certainly have new equilibrium points 
for both  $0<\mu<0.5$ and $\mu=0.5$ cases.

{\bf  Corollary {3.6}}\\ {\it If 
$T<\sqrt{2}\mu$ and
 $Q_2(-T/\sqrt{2})>P_2(-T/\sqrt{2})$, 
then we will have new equilibrium points. 
That is, there are at least five equilibrium points on the $x$-axis
for System (\ref{eq:ma3body2}).}

One can see that the conditions studied in Property 3.4 and 3.5 do
not include all possible cases. That is, there are some conditions we fail
to provide analytical results due to that the results of those cases would
more strongly depend on the detail structure of the function $P_2(x)$ and
$Q_2(x)$. Thus,  Corollary {3.6} shows that the condition, 
$T<\sqrt{2}\mu$ and $Q_2(-T/\sqrt{2})>P_2(-T/\sqrt{2})$, is only a sufficient
condition to have new equilibrium points. It is not a necessary condition.




To check the above statements, 
we numerically solve both Eq.(\ref{eq:fxe}) and Eq. (\ref{eq:gxe}) 
and  find out the number of 
equilibrium points for different given parameters.
The numerical scheme of root finding 
is the Van Wijngaarden-Dekker-Brent Method (Brent 1973). This is 
an excellent algorithm recommended by Press et al. (1992).
We set a high level of  
accuracy that the maximum error is  $10^{-8}$ for the locations of 
equilibrium points on both $x$-axis and the line: $x=1/2-\mu$. 
All the new equilibrium points are
on the $x$-axis as we have proved in the analytic results.
Figure 1 is the results on the $\mu-M_b$ plane.
That is, we numerically search the solutions of both 
Eq.(\ref{eq:fxe}) and Eq.(\ref{eq:gxe}) for different $(\mu, M_b)$,
where $\mu$ is between [0,0.5] 
and $M_b$ is between [0, 0.6].
Those $(\mu, M_b)$ with new equilibrium points are marked by points.

In each panel, there is a solid curve which satisfies
$Q_2(-T/\sqrt{2})=P_2(-T/\sqrt{2})$ with given parameters.
Figure 1(a) shows that this curve perfectly match the boundary of the
existence area of new equilibrium points. Therefore, 
 Corollary 3.6 precisely predicts our numerical results. 
However, as we mentioned earlier, our analytic results do not include
all possible conditions.
Thus, the solid curve and the boundary of the existence 
area of new equilibrium
points do not match that well in Figure 1(b)-(d). 
We cannot see the solid curve in Figure 1(d)
since this curve is out of our studied parameter space. 

Moreover, 
Figure 1 shows that the area without any points are on the left-bottom 
of $\mu-M_b$ plane.  This is reasonable since 
(1) when $M_b$ is small, it becomes classical restricted three-body problem
and (2)
when $\mu$ is small, this system become similar to a planetary system
with a belt.

From Figure 1(a), i.e. $T=0.01$ case, 
it is interesting that there is a lower limit of mass ratio:
below that ratio, there would
be no new equilibrium point
no matter how large the belt mass is. 
Such limits of mass ratio become larger for larger $T$ as in Figure (b)-(d).
In general, when the mass of the belt is larger, 
it is easier to have new equilibrium points for larger mass ratio.
It is interesting that the new equilibrium points 
could exist for the larger area  
on the $\mu-M_b$ plane 
when $T$ is smaller. 
It is also interesting that 
the locations and number of equilibrium points depend on $T$ but not 
directly on $a$ and $b$.  
$T$ is the summation of $a$ 
(the flatness parameter) and $b$ (the core parameter) and 
the Miyamoto-Nagai potential would be equivalent to 
the potential of a point mass when $T=0$.

Figure 2 is the results on the $T-M_b$ plane.
That is, we numerically search the solutions of both 
Eq.(\ref{eq:fxe}) and Eq.(\ref{eq:gxe}) for different $(T, M_b)$,
where $T$ is between [0,0.4] with grid size 0.01
and $M_b$ is between [0, 0.6] with grid size 0.02. 
Those $(T, M_b)$ with new equilibrium points are marked by points.

Figure 2 shows that when $T$ is small enough, 
we always have new equilibrium points for any value of belt mass. We also 
plot out the solid curve: $Q_2(-T/\sqrt{2})=P_2(-T/\sqrt{2})$ in each panel. 
The left-up side of the solid curve satisfies the condition
of Corollary 3.6, i.e. $Q_2(-T/\sqrt{2})>P_2(-T/\sqrt{2})$,
so we have new equilibrium points in this area.
We find that these solid curves can match the boundaries of the 
existence area of new
equilibrium points very well for Figure 2(a)-(c) but not that well for 
Figure 2(d). Nevertheless, they all indicate that our analytic results
are completely consistent with our numerical results. 

In addition to the full circle points, 
for the comparison purpose here,
there are also open triangle points
in Figure 2(d). The open triangle points are the results for 
$n=\sqrt{1-2f_b(0.5)}$ as in Jiang \& Yeh (2006). 
The open triangle points almost cover the same area as the one covered
by the full circle points, thus our simplification that $n=1$ is a good
approximation.

\subsection{Model B}

On the other hand, we also study Model B: Power-Law Profile. 
By Property 3.1, the equilibrium points $(x_e, y_e)$ must be on the lines: 
$x=1/2-\mu$ or $y=0$. Property 3.1 and 3.2 indicates that 
 we only need to solve $x_e$ from
\beq
k(x)= x-\frac{\mu_1(x+\mu_2)}{r_1^3}-\frac{\mu_2(x-\mu_1)}
{r_2^3}-\frac{2x}{r^2}  \int^{r_o}_{r_i}\rho(r')r'
\left[\frac{E}{r-r'}+\frac{F}{r+r'}\right]dr'\biggm|_{y=0}=0 
\eeq
and $y_e$ from
\beq
h(y)=1 - \left[\frac{1}{4}+y^2\right]^{-3/2} 
-\frac{2}{r^2}  \int^{r_o}_{r_i}\rho(r')r'\left[\frac{E}{r-r'}
+\frac{F}{r+r'}\right]dr'\biggm|_{x=(\mu_1-\mu_2)/2} =0 
\eeq
to get all the equilibrium points $(x_e, 0)$ and 
$(1/2-\mu, y_e)$. Because there are Elliptic Integrals in
$k(x)$ and $h(y)$, it is very difficult to get any analytic results 
as Property 3.3, 3.4 and 3.5 of Model A. Thus, 
we try to find the roots of $k(x)$ and $h(y)$ numerically  
and we only present the numerical results here.
The numerical method is exactly the same as the one used in Model A.

Contrary to Model A,
we find that the new equilibrium points can exist both on the vertical line: 
$x=1/2-\mu$ and on the $x$-axis. 
Figure 3 and Figure 4 are the results, where the cross points are those
area when there are more than three roots for $k(x)=0$,
the circles are the area when there are more than two roots for
$h(y)=0$. Please note that
for the classical restricted three-body problem, there
are three collinear points and two 
triangular ones.

Figure 3 is the results on the $\mu-M_b$ plane.
That is, we numerically search the solutions of both 
Eq.(\ref{eq:fxe}) and Eq.(\ref{eq:gxe}) for different $(\mu, M_b)$,
where $\mu$ is between [0,0.5]
and $M_b$ is between [0, 0.6].

The area of circles in 
Figure 3(a) looks  like a square, thus there is a lower limit
of mass ratio that no matter how large the belt mass is, there would be 
no new equilibrium point. 
There is also a lower limit of belt mass that no matter 
what mass ratio is, there would be no new equilibrium point.
Figure 3(b) and (c) are very similar but the area becomes 
more like a quarter of a circle.
The existence area of new equilibrium points 
in Figure 3(d) is very different from  
others. 

The cross points in Figure 3 only appear for a small range of mass ratio.
That is, with the additional force from the belt, the mass ratio
has to be a particular value to have new equilibrium points.
This means that it is more difficult to have new equilibrium points 
on the $x$-axis than that on the $y$-axis.
To have new equilibrium points on the $y$-axis, the centrifugal force
has to balance with the $y$-component of the total force from 
both central binary objects plus the force from the belt.
That is, there are three components to balance each other at the 
equilibrium points and the mass ratio would only modify the 
locations of equilibrium points.
However, on the $x$-axis, the forces from two central binary objects
are in different directions at new equilibrium points, thus 
there are four components to balance each other for this case. 
Therefore, it will be more difficult for them to get balanced. 
To have new equilibrium points, 
the mass ratio has to be within a particular range.

Figure 4 is the results on the $r_i-M_b$ plane.
That is, we numerically search the solutions of both 
Eq.(\ref{eq:fxe}) and Eq.(\ref{eq:gxe}) for different $(r_i, M_b)$,
where $r_i$ is between [0,1] with grid size 0.01
and $M_b$ is between [0, 0.6] with grid size 0.02. 

The cross points in Figure 4 show that 
there is a gap of $r_i$ where it is more difficult to have 
new equilibrium points on the $x$-axis. 
It seems that this gap become larger when $\mu$
is larger.
The reason why it would be more difficult to have new  equilibrium points
for particular region of $r_i$
might be related to the relative positions of 
central binary objects with respect to the belt.
The number of separate components which could balance each other 
might influence the existence of new equilibrium points.

The circles in Figure 4 are very interesting. 
In Figure 4(a) where the mass ratio is $\mu=1/11$, 
there is a triangle area, i.e.
when $0.5 < r_i < 0.8$, the range of belt mass to 
have new equilibrium points is larger if $r_i$ is larger.
However, 
when $r_i < 0.5$ or  $r_i > 0.8 $,  there is no new equilibrium point.
The triangle becomes slightly larger in Figure 4(b) and 
it becomes the combination of two trapezoids facing each other
in Figures 4(c)-(d).

\section{Stability Analysis}

When the locations of equilibrium points are determined, it would be 
interesting to understand the stability properties around these points.
We now consider the following system: 
\beq \left\{
\begin{array}{ll}
&  \frac{dx}{dt}=u  \\
& \frac{dy}{dt}=v  \\
& \frac{du}{dt}=2v +f(x,y)\\
& \frac{d v}{dt}=-2u+g(x,y),
\end{array}\right. 
\eeq
where $f(x,y)$, $g(x,y)$ are defined in Eqs.(\ref{eq:gf2})-(\ref{eq:gg2}).
Following the usual linearization, the eigenvalues $\lambda$ corresponding
to 
the equilibrium points would satisfy: 
\beq
\lambda^4+(4-f_x-g_y)\lambda^2+2(f_y-g_x)\lambda+f_xg_y-g_xf_y=0, 
\label{eq:labd1}
\eeq
where $f_x\equiv {\partial f(x,y)}/{\partial x}$ and $f_y, g_x, g_y$
are also defined similarly.

Consider a new equilibrium point $(x_e, y_e)$ of Model A as an example, so 
$(x_e,y_e)$ satisfies $k(x_e)=0$ and $y_e=0$. Thus, we have:
\beqn
& &f_x(x_e,0)=\left(\frac{\mu}{x_e}+3\right)\frac{1-\mu}{|x_e+\mu|^3}
-\left(\frac{1-\mu}{x_e}+3\right)\frac{\mu}{|x_e+\mu-1|^3}+\frac{3M_bx_e^2}
{(x_e^2+T^2)^{5/2}}, \non \\
&& g_y(x_e,0)=\frac{\mu(1-\mu)}{x_e}\left(\frac{1}{|{x_e}-\mu|^3}
-\frac{1}{|{x_e}+\mu-1|^3}
\right).
\eeqn

Since $f_y(x_e,0)=0$ and $g_x(x_e,0)=0$,  Eq.(\ref{eq:labd1}) becomes
\beq
\lambda^4+(4-f_x-g_y)\lambda^2+f_x g_y=0. 
\label{eq:labd2}
\eeq
For convenience, we define $\Omega=f_xg_y$ and
$\Pi=f_x+g_y-4$. Therefore, we have
\beq
\lambda^2_{+}=\frac{\Pi+\sqrt{\Pi^2-4\Omega}}{2} \quad {\rm and} \quad 
\lambda^2_{-}=\frac{\Pi-\sqrt{\Pi^2-4\Omega}}{2}.
\label{eq:lamda3}
\eeq
These two relations will determine
the eigenvalues and the properties of the equilibrium point.
In fact, the signs of $\Pi^2-4\Omega$, $\Pi+\sqrt{\Pi^2-4\Omega}$, and
$\Pi-\sqrt{\Pi^2-4\Omega}$ will completely determine the results.
However, when one knows the signs of $\Pi$ and $\Omega$, the details
of possible eigenvalues can be directly determined by them.
We therefore list all the combinations in Table 1,
in terms of the signs of $\Pi^2-4\Omega$, $\Pi$ and $\Omega$.

\vskip 0.2truein
 {\small \centerline{ {\bf Table 1.} The Possible Combinations}
    \begin{center}
       \begin{tabular}{|l|l|l|l|c|c|}\hline
Condition A & Condition B & Condition C & Row &$\lambda^2_{+}$&
$\lambda^2_{-}$\\\hline
$\Pi^2-4\Omega<0$&      &      &  1 & c & c          \\\hline
$\Pi^2-4\Omega=0$& $\Pi<0$&  &   2  & $-$ & $-$ \\\hline
$\Pi^2-4\Omega=0$& $\Pi=0$&  &   3  & 0 & 0\\\hline
$\Pi^2-4\Omega=0$& $\Pi>0$&  &    4 & + & +\\\hline
$\Pi^2-4\Omega>0$& $\Pi<0$& $\Omega<0$&   5 & + & $-$  \\\hline
$\Pi^2-4\Omega>0$& $\Pi<0$& $\Omega=0$&   6 & 0 & $-$ \\\hline
$\Pi^2-4\Omega>0$& $\Pi<0$& $\Omega>0$&   7 & $-$ & $-$  \\\hline
$\Pi^2-4\Omega>0$& $\Pi=0$& $\Omega<0$&   8 & + & $-$ \\\hline
$\Pi^2-4\Omega>0$& $\Pi=0$& $\Omega=0$&   9 (impossible) & &   \\\hline
$\Pi^2-4\Omega>0$& $\Pi=0$& $\Omega>0$&   10 (impossible) & &  \\\hline
$\Pi^2-4\Omega>0$& $\Pi>0$& $\Omega<0$&   11 & + & $-$   \\\hline
$\Pi^2-4\Omega>0$& $\Pi>0$& $\Omega=0$&   12 & + & 0 \\\hline
$\Pi^2-4\Omega>0$& $\Pi>0$& $\Omega>0$&   13 & + & $+$  \\\hline
\end{tabular}
    \end{center}
    }
\vskip 0.2truein

 \normalsize


In Row 1, due to $\Pi^2-4\Omega<0$, $\lambda^2_{\pm}$ are then imaginary
complex conjugate numbers. Thus, $\lambda_{\pm}$ can be expressed as 
$\lambda_{+}=\pm a\pm b {\rm i}$ and 
$\lambda_{-}=\pm c\pm d{\rm i}$, where $a$,$b$,$c$ and $d$ are positive 
real numbers.
So in this case, the equilibrium point is an unstable point.

In Row 2, 3, and 4,
$$\lambda^2_{+}=\lambda^2_{-}=\frac{\Pi}{2}.$$
Thus, $\lambda_{\pm}$ could be pure imaginary (Row 2), zeroes (Row 3) and 
$\pm\sqrt{\frac{\Pi}{2}}$ (Row 4). The equilibrium point is 
neutrally stable for Row 2 and 3 but unstable for Row 4.

In Row 5, because $\Omega < 0$, we have $\Pi^2-4\Omega > \Pi^2$. 
Thus, $\sqrt{\Pi^2-4\Omega} > |\Pi| = -\Pi$. In this case,
 $\Pi+\sqrt{\Pi^2-4\Omega}>0$ and $\Pi-\sqrt{\Pi^2-4\Omega}<0$.
We then know that $\lambda_{+}=\pm a$, where $a$ is a positive real number;
$\lambda_{-}=\pm d{\rm i}$, where $d$ is a positive real number.
So, the equilibrium point is an unstable point.

In Row 6, because $\Omega = 0$, we have 
$$\lambda^2_{\pm}=\frac{\Pi\pm|\Pi|}{2}.$$
Because $\Pi<0$, $\lambda^2_{+}=0$ and $\lambda^2_{-}=\Pi<0$.
Thus, $\lambda_{+}=0$ and $\lambda_{-}$ is pure imaginary.
So in this case, the equilibrium point is a neutrally stable point.

In Row 7, because $\Omega > 0$, we have 
$\Pi^2-4\Omega<\Pi^2$. Therefore, $\sqrt{\Pi^2-4\Omega}<|\Pi|=-\Pi$.
In this case,
$\Pi+\sqrt{\Pi^2-4\Omega}<0$ and $\Pi-\sqrt{\Pi^2-4\Omega}<0$.
Both $\lambda_{+}$ and $\lambda_{-}$ are pure imaginary.
So, the equilibrium point is a neutrally stable point.

In Row 8, because $\Pi=0$, we have 
$$\lambda^2_{+}=\frac{\sqrt{-4\Omega}}{2}  \quad {\rm and} \quad
 \lambda^2_{-}=\frac{-\sqrt{-4\Omega}}{2}.$$
Since $\Omega<0$, thus  $\lambda^2_{+}>0$ and  $\lambda^2_{-}<0$.
We then have that $\lambda_{+}=\pm a$, where $a$ is a positive real number;
$\lambda_{-}=\pm d{\rm i}$, where $d$ is a positive real number.
So, the equilibrium point is an unstable point.

In Row 9 and 10, Condition C contradicts with Condition A after we use
Condition B: $\Pi=0$. Thus, these two are impossible.

In Row 11, because $\Omega < 0$, we have $\Pi^2-4\Omega > \Pi^2$. 
Thus, $\sqrt{\Pi^2-4\Omega} > |\Pi| = \Pi$. In this case,
 $\Pi+\sqrt{\Pi^2-4\Omega}>0$ and $\Pi-\sqrt{\Pi^2-4\Omega}<0$.
We then know that $\lambda_{+}=\pm a$, where $a$ is a positive real number;
$\lambda_{-}=\pm d{\rm i}$, where $d$ is a positive real number.
So, the equilibrium point is an unstable point.

In Row 12, because $\Omega = 0$, we have 
$$\lambda^2_{\pm}=\frac{\Pi\pm|\Pi|}{2}.$$
Because $\Pi>0$, $\lambda^2_{+}=\Pi$ and $\lambda^2_{-}=0$.
Thus, $\lambda_{+}=\pm\sqrt{\Pi}$ 
and $\lambda_{-}=0$.
So in this case, the equilibrium point is unstable.

In Row 13, because $\Omega > 0$, we have 
$\Pi^2-4\Omega<\Pi^2$. Therefore, $\sqrt{\Pi^2-4\Omega}<|\Pi|=\Pi$.
In this case,
$\Pi+\sqrt{\Pi^2-4\Omega}>0$ and $\Pi-\sqrt{\Pi^2-4\Omega}>0$.
Both $\lambda_{+}$ and $\lambda_{-}$ are real 
and one of their roots is positive.
So, the equilibrium point is unstable.

In general, from the above analysis, there are two categories: 
(i) when all the four eigenvalues' real parts are zero, the
equilibrium point is neutrally stable, (ii) as long as one of the eigenvalues 
has a positive real part, the equilibrium point is unstable.
For this system, it is impossible that all the four eigenvalues' real parts 
are negative. In Table 1, Row 2, 3, 6, 7 belong to the first category
because their eigenvalues are either zero or pure imaginary. Row 1, 4, 5, 8,
11, 12, 13
belong to the second category because at least one of the eigenvalues is
positive.

To make it convenient to use Table 1 in the future, the properties of
$\lambda_{+}^2$ and $\lambda_{-}^2$ are indicated in the final two columns
of the table. In that two columns, ``c'' means imaginary
complex conjugate numbers, ``+'' means positive numbers, ``$-$''
means negative numbers, and ``0'' means zeroes. 
 

   


As an example, when we set $M_b=T=0.01$ and $\mu=4/9$ in Model A, 
there are five equilibrium points on $x$-axis:
(-1.180392,0), (-0.060183,0), (-0.000137,0), (0.118920,0), (1.218591,0).
We find that only (-0.000137,0) belongs to the first category
and is neutrally stable. All the others belong to the second category
and is unstable.

\section{Concluding Remarks}

We have studied the existence of new equilibrium points for 
our system. In addition to the usual  
equilibrium
points, we find that there are some new equilibrium points for particular 
parameter space.
We provide two models in this paper. One is an analytic model from
Miyamoto \& Nagai (1975) (Model A) 
and another is a model of power-law density profile (Model B).
Since there is no analytic formula for the power-law density profile,
we have to study this model numerically. 
For Model A, we have  some interesting analytic results
and they are consistent with the numerical results. 
The new equilibrium points only exist on the $x$-axis for Model A.
However, for Model B, we can have new equilibrium points
 both on the vertical line:
$x=1/2-\mu$ and the $x$-axis. 
 
Our study implies that the structure of constant potential contour
might be more complicated due to the existence of new equilibrium points.
The stability could be different according to the linear analysis
in the previous section.  
Thus, the dynamics around a binary system might become very different when the 
influence of a belt is included. 

\section*{Acknowledgment}

We are grateful to the National Center for High-performance Computing
for computer time and facilities. 
This work is supported in part 
by the National Science Council, Taiwan, under Li-Chin Yeh's 
Grants NSC 94-2115-M-134-002 and also Ing-Guey Jiang's
Grants NSC 94-2112-M-008-010.



\begin{figure}[tbhp]
\epsfysize 7.0 in \epsffile{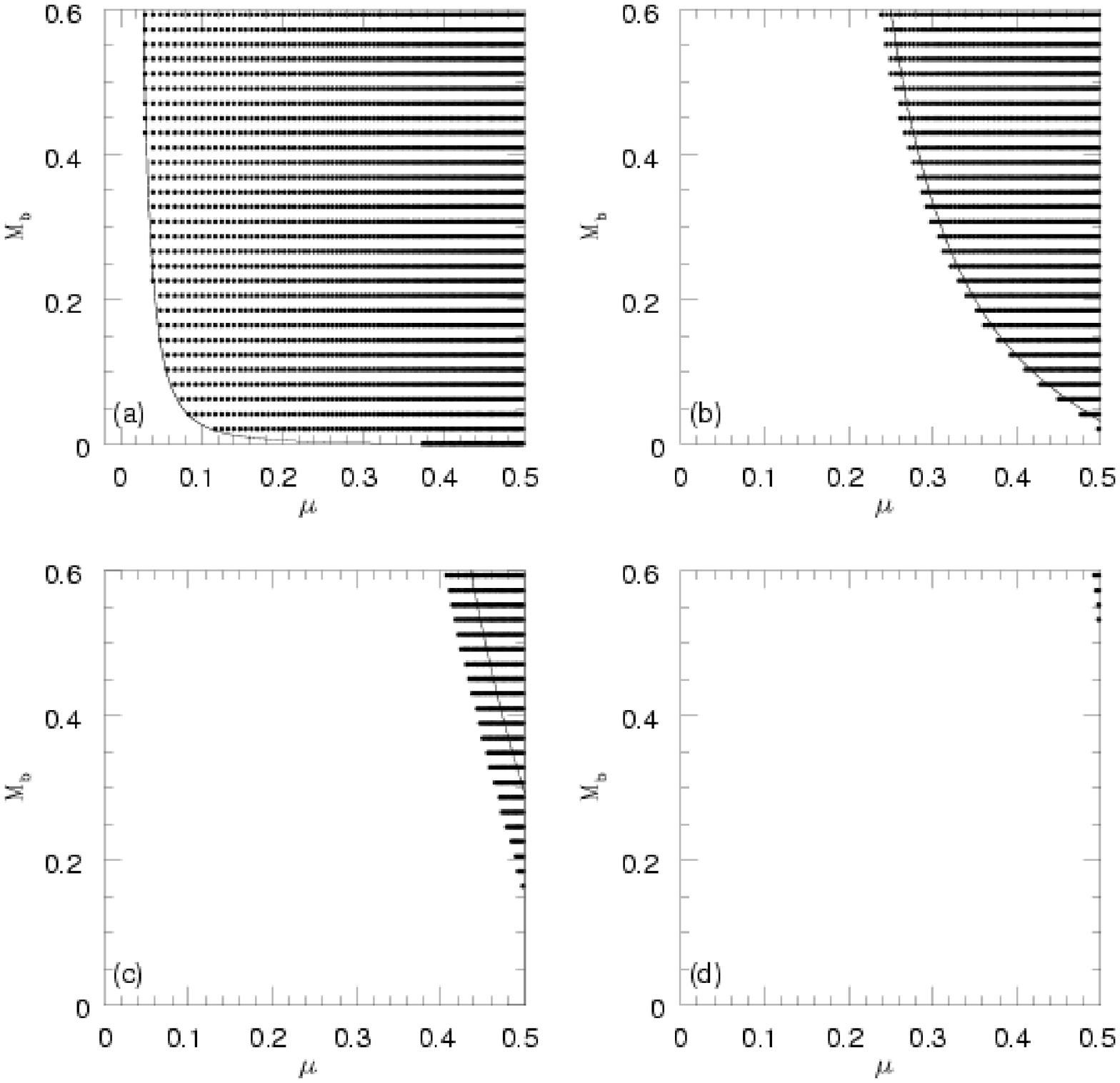}
\caption{The existence area of new equilibrium points on the
$\mu-M_b$ plane for Model A. 
In (a) $T=0.01$; 
(b) $T=0.1$; (c)$T=0.2$ and (d) $T=0.3$
(please see the main text for more details).}
\end{figure}

\clearpage

\begin{figure}[tbhp]
\epsfysize 7.0 in \epsffile{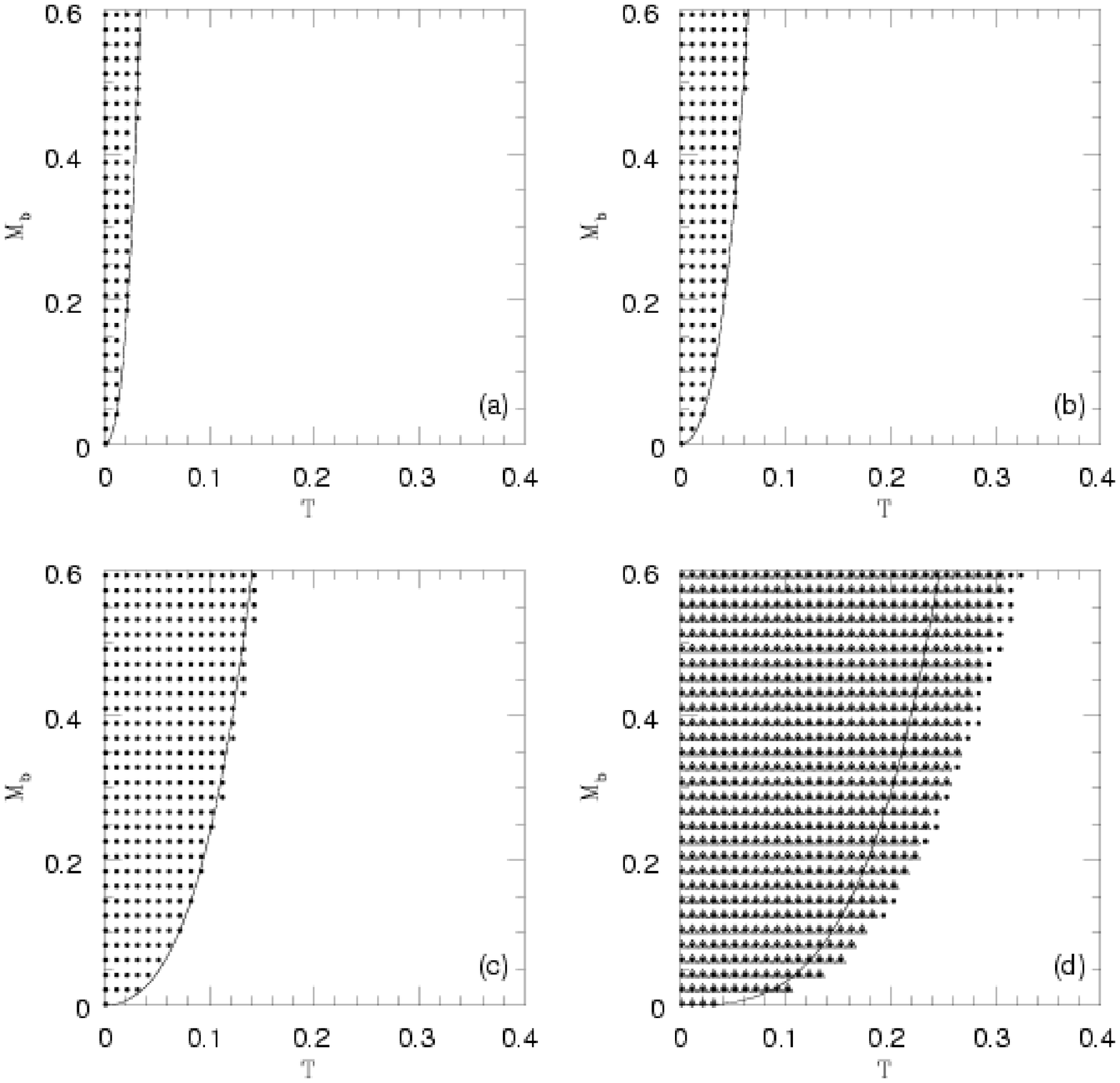}
\caption{The existence area of new equilibrium points on the
$T-M_b$ plane for Model A. 
In (a) $\mu=1/11$; (b) $\mu=1/6$; (c)$\mu=1/3$ and (d) 
$\mu=1/2$
(please see the main text for more details).}
\end{figure}

\clearpage

\begin{figure}[tbhp]
\epsfysize 7.0 in \epsffile{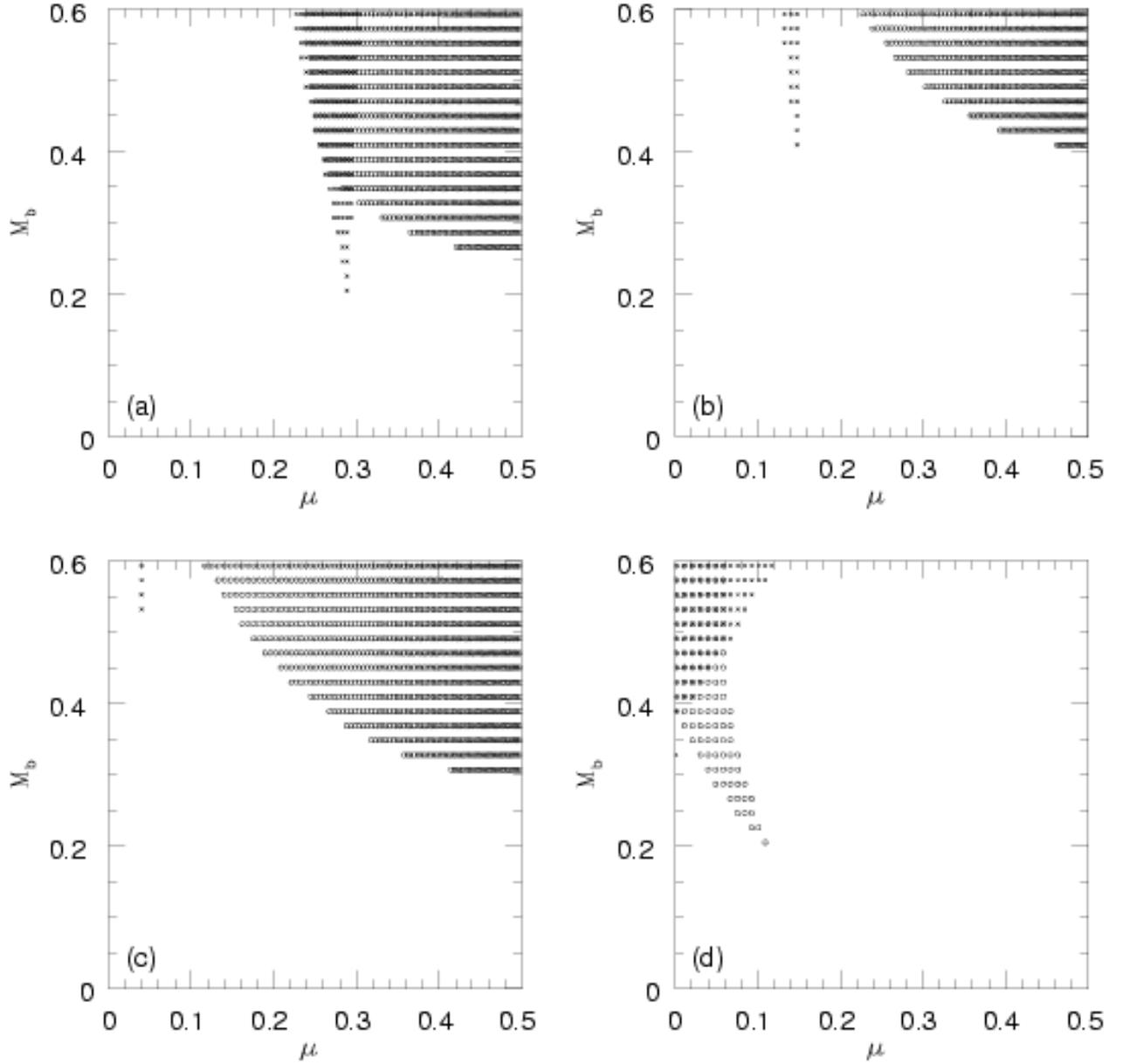}
\caption{The existence area of new equilibrium points on the
$\mu-M_b$ plane for Model B. 
In (a) $r_i=0.2$; 
(b) $r_i=0.4$; (c)$r_i=0.6$ and (d) $r_i=0.8$
(please see the main text for more details).}
\end{figure}

\clearpage

\begin{figure}[tbhp]
\epsfysize 7.0 in \epsffile{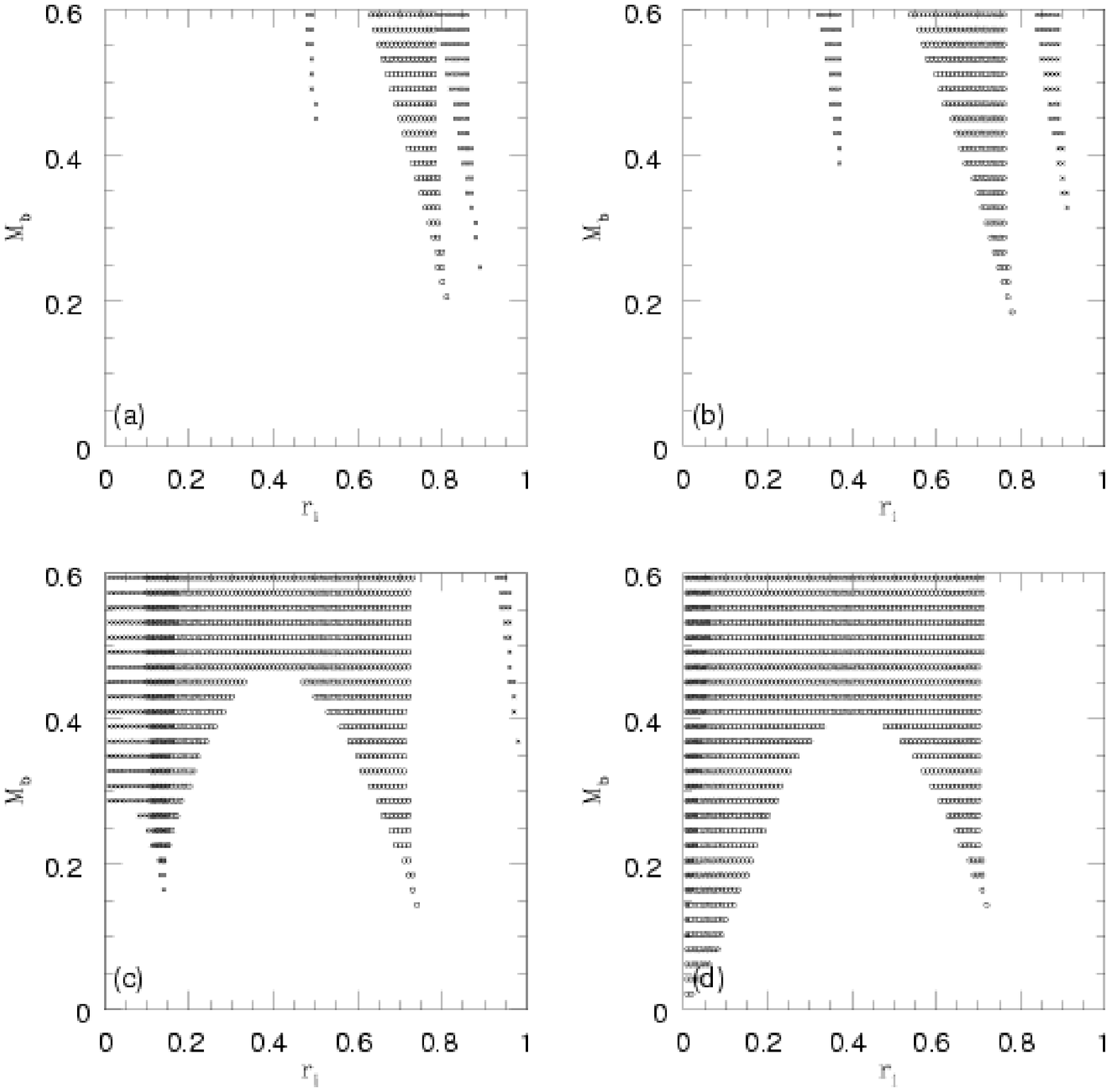}
\caption{The  existence area of new equilibrium points on the
$r_i-M_b$ plane for Model B.
In (a) $\mu=1/11$; (b) $\mu=1/6$; (c)$\mu=1/3$ and (d) 
$\mu=1/2$ (please see the main text for more details).}
\end{figure}

\end{document}